\documentclass[aps,amsmath,amssymb,twocolumn,superscriptaddress,nofootinbib]{revtex4-2}

\usepackage{graphicx}
\usepackage{dcolumn}
\usepackage{bm}
\usepackage{color}
\usepackage[colorlinks=true,pdfstartview=FitV,linkcolor=blue,citecolor=blue,urlcolor=blue]{hyperref}
\usepackage[mathlines]{lineno}
\usepackage[dvipsnames]{xcolor}
\usepackage[utf8]{inputenc}
\usepackage[T1]{fontenc}
\usepackage{newtxtext,newtxmath}
\usepackage[dvipsnames]{xcolor}
\usepackage{multirow}
\usepackage{threeparttable}
\usepackage{booktabs}
\usepackage{float}
\usepackage{amsmath}
\usepackage{gensymb}
\everymath{\displaystyle}

\usepackage{enumitem}
\setlist[enumerate]{label*=\arabic*.}

\usepackage{xcolor}

\begin{document}

\title{Demonstration of a Quantum Noise Limited Traveling-Wave Parametric Amplifier}

\author{Nikita~Klimovich}\email{nikita@alumni.caltech.edu}\affiliation{Division of Physics, Mathematics and Astronomy, California Institute of Technology, Pasadena, CA 91125, USA} 
\author{Peter~K.~Day}\affiliation{Jet Propulsion Laboratory, California Institute of Technology, Pasadena, CA 91109, USA}
\author{Shibo~Shu}\affiliation{Division of Physics, Mathematics and Astronomy, California Institute of Technology, Pasadena, CA 91125, USA}
\author{Byeong~Ho~Eom}\affiliation{Jet Propulsion Laboratory, California Institute of Technology, Pasadena, CA 91109, USA} 
\author{Henry~G.~LeDuc}\affiliation{Jet Propulsion Laboratory, California Institute of Technology, Pasadena, CA 91109, USA}
\author{Andrew~D.~Beyer}\affiliation{Jet Propulsion Laboratory, California Institute of Technology, Pasadena, CA 91109, USA}

\date{\today}

\begin{abstract}
Recent progress in quantum computing and the development of novel detector technologies for astrophysics is driving the need for high-gain, broadband, and quantum-limited amplifiers. We present a purely traveling-wave parametric amplifier (TWPA) using an inverted NbTiN microstrip and amorphous Silicon dielectric. Through dispersion engineering, we are able to obtain $50~\Omega$ impedance matching and suppress undesired parametric processes while phase matching the three-wave-mixing amplification across a large range of frequencies. The result is a broadband amplifier operating with 20dB gain and quantum-limited noise performance at 20mK. At the single frequency where the amplifier is phase sensitive, we further demonstrate 8dB of vacuum noise squeezing.
\end{abstract}

\maketitle


Experiments operated at microwave frequencies in need of low-noise amplification typically rely on semiconductor high-electron mobility transistor (HEMT) amplifiers. Although HEMTs provide excellent gain and bandwidth, their noise temperatures of a few Kelvin\cite{osti_1471452} are consistently an order of magnitude above the theoretical quantum limit\cite{Caves}. If bandwidth is not an issue, Josephson parametric amplifiers (JPAs) could provide quantum-limited noise performance,\cite{PhysRevA.39.2519,Tholen} but the bandwidth of such devices was fundamentally limited by the resonant cavity used to provide sufficient interaction length with the nonlinear junction.\cite{JPA_Beltran}

The past decade has seen two major breakthroughs of parametric amplifier technology. Improved fabrication techniques have allowed the creation of thousands of Josephson junctions in series, enabling the production of broadband JPAs.\cite{JPA_Macklin} JPAs have been used to demonstrate broadband 20 dB gain, quantum-limited noise, and vacuum noise squeezing beyond the quantum limit.\cite{JPA_Qiu_2023} While this monumental progress and potential uses of these devices should not be understated, JPAs are not necessarily suitable for all experiments in need of quantum-limited amplification due to the relatively small dynamic range limited by the critical current of the junctions.\cite{doi:10.1063/5.0064892}

In parallel with these developments, it was shown that some superconducting materials can provide a sufficient kinetic inductance nonlinearity for a purely traveling wave parametric amplifier (TWPA) given proper dispersion engineering.\cite{JDE} The kinetic inductance of a superconducting film will increase nonlinearly due to the change the density of states that results from an applied current\cite{Anthore_2003} and can be exploited to produce wave-mixing that can be used to create parametric amplifiers, frequency converters, and other parametric devices.\cite{BOYD200869} The main advantages of TWPAs are twofold: first, such devices are easier to fabricate as they only require a uniform conducting layer in comparison to the thousands of working junctions of JPAs; second, the lack of junctions substantially increases the critical current, which allows for the amplification of stronger signals before reaching the 1 dB saturation point. 

\begin{figure*}[!htp]
    \centering
    \includegraphics[width=0.95\textwidth]{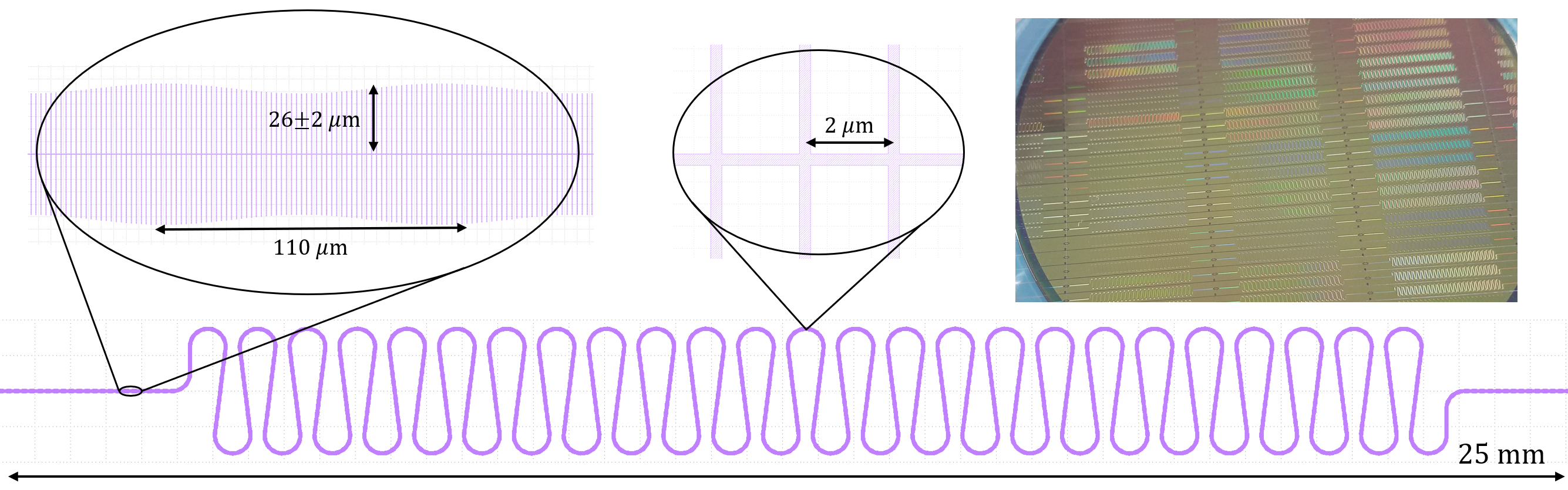}
    \caption{A top view of the TWPA design and a photo of 76 such devices on a four inch wafer \textit{(top-right)}.}
    \label{fig:ParampFigure}
\end{figure*}  

Since the initial demonstration of a kinetic inductance TWPA operating on these principles,\cite{JDE} the need for robust, broadband, quantum-limited amplification has driven the development of many similar devices.\cite{PRXQuantum.2.010302,10.1063/1.5063252,10.1063/1.4937922,Bockstiegel,10.1063/1.4980102,7456219,10.1063/5.0004236,Shu_2021} Despite the excellent progress made by these groups, obtaining a substantial broadband 20 dB gain simultaneously with quantum-limited noise performance has proved elusive. In this paper, we present an experimental demonstration of such a TWPA and further use it to generate a squeezed vacuum state. 

Our TWPA is a kinetic-inductance-based traveling-wave device consisting of an inverted microstrip fabricated on top of a Silicon wafer. The transmission line is composed of a 35 nm thick, 250 nm wide superconducting NbTiN ($T_c = 14.5$K; $\rho = 200\mu\Omega$cm). The length of the transmission line is 86~mm meandered to fit within a 26 mm by 5 mm footprint. The microstrip conductor is separated from the 200 nm thick NbTiN sky plane by a 190 nm amorphous Silicon dielectric. The transmission line design is low loss\cite{Shu_2021} and robust with respect to short-to-ground defects that can afflict coplanar transmission line geometries.

To increase the capacitance per unit length and obtain 50$\Omega$ impedance, 250 nm wide, 26 $\mu$m long open stub sections are placed periodically on either side the central line with a separation of 2 $\mu$m (Figure \ref{fig:ParampFigure}). The large kinetic inductance and capacitance of this geometry results in a very low phase velocity of $v_\text{ph} = 0.0078c$. These capacitive stubs also act as 86 thousand quarter-wave resonators near 150 GHz, which results in a large nonlinearity in the dispersion curve at frequencies above the pump, altering the phase matching criterion of parametric processes at higher frequencies and thereby suppressing the formation of pump harmonics.\cite{klimovich2022traveling} The average length of these capacitive stubs is modulated by a sinusoid of 2 $\mu$m amplitude and 110 $\mu$m wavelength to introduce a periodic modulation in the impedance that results in a photonic band gap near 9 GHz for optimally phase matching the three-wave-mixing gain. 

\begin{figure}[!htp]
    \centering
    \includegraphics[width=0.48\textwidth]{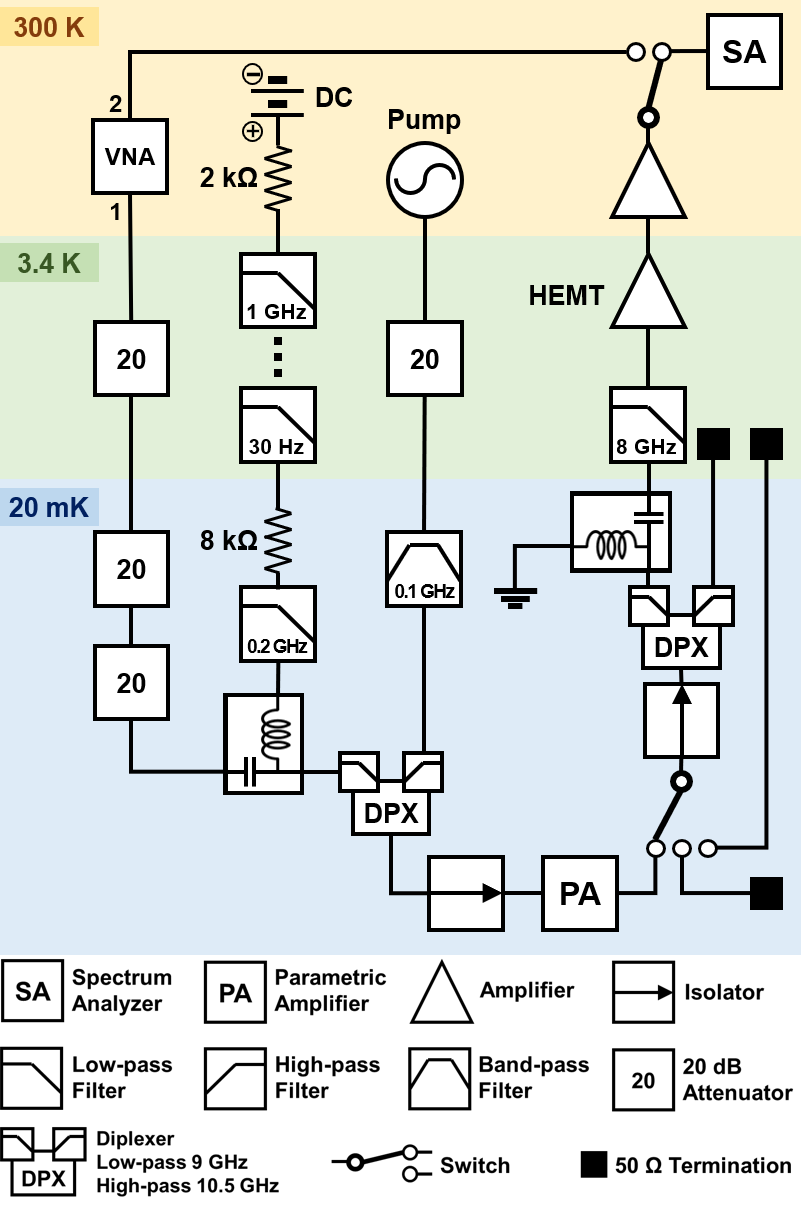}
    \caption{A schematic diagram of the measurement setup.}
    \label{fig:NoiseSetup}
\end{figure}

We characterize the gain and noise performance of the amplifier by using a modified y-factor measurement setup shown in Figure \ref{fig:NoiseSetup}. The input line from the VNA is strongly attenuated to reduce the room-temperature thermal noise contributions at the TWPA input to a 1 mK level, and is then combined with the filtered DC and pump lines. The output of the system is amplified in two-stages using a HEMT at 4 K and room-temperature amplifier. To avoid gain saturation of these post-amplifiers, the pump is separated using a diplexer and further attenuated by a low-pass filter. A cryogenic switch changes between the TWPA and known hot/cold loads for calibrating the spectrum analyser signal to noise temperature at the TWPA's output. 

The combination of these components introduces additional reflections into the system. If the gain of the TWPA exceeds the losses from the round-trip reflections at the input and output of the device, the system will undergo parametric oscillation. Once this occurs, a signal at the TWPA input will grow exponentially until pump depletion lowers the overall gain of the device beneath the aforementioned criterion.\cite{klimovich2022traveling} This oscillation can be a substantial limit on TWPA performance and will lower the gain by nearly 10 dB if the reflections from the system components are unaddressed. These reflections are eliminated by a pair of isolators immediately preceding and following the TWPA. An exception is made for the cryogenic switch connecting the cold loads, which is placed before the output isolator to avoid any systematics associated with non-common-path components during our noise measurement. This concession results in 2 dB lower achievable gain from the TWPA compared to an identical setup with the switch placed after the isolator.

\begin{figure}[!htp]
    \centering
    \includegraphics[width=0.48\textwidth]{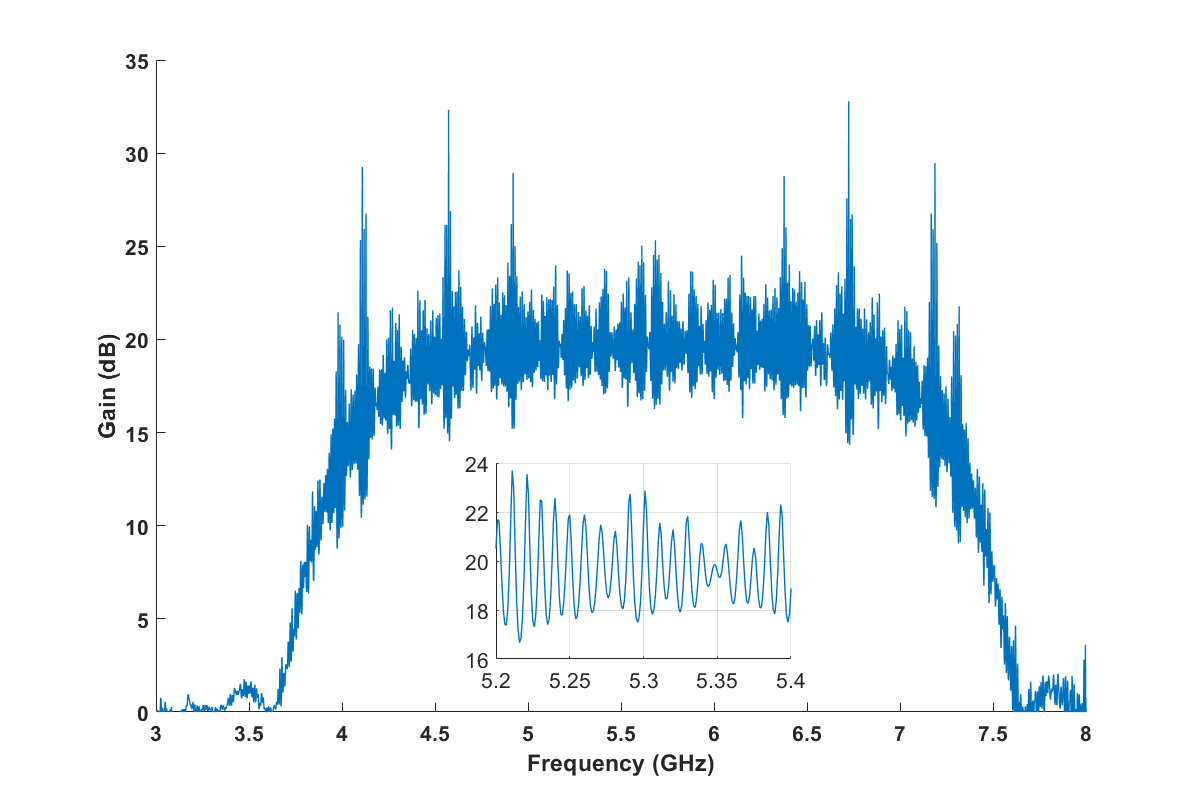}
    \caption{The three-wave-mixing parametric amplifier gain measured at 20 mK in the setup shown in Figure \ref{fig:NoiseSetup} from an 11.297 GHz -17.2 dBm pump (at the signal generator) and 0.579 mA DC current. \textit{Inset:} A zoom-in showing the scale of the gain ripple.}
    \label{fig:AddedNoiseGain}
\end{figure}

A measurement of the the relative on/off gain of the TWPA within this setup is shown in Figure \ref{fig:AddedNoiseGain}. We obtain a broadband 20 dB gain across 2GHz of bandwidth with 15 dB gain over more than 3GHz of frequency range. The losses within our films were carefully measured in a previous experiment to be $0.038$ dB/GHz.\cite{Shu_2021} Note that the change in the impedance, $L$, from the applied DC current alters both the phase velocity ($v_\text{ph} = 1/\sqrt{\mathcal{LC}}$) and impedance ($Z = \sqrt{\mathcal{L}/\mathcal{C}}$) of the line, changing the amplitude and frequency of the standing wave ripples of the $S_{21}$. To account for this effect and better represent the true gain of the device we have chosen to present and the amplifier-on case relative to the DC-on (no applied pump) rather than both pump and DC off. This choice noticeably reduces the systematic fluctuations in our plots corresponding with the primary 8~MHz-spaced ripple in our $S_{21}$ but does not otherwise impact the results. In principle, a similar effect is introduced by the nonlinearity due to the applied pump tone, but the resulting shift in ripple frequencies is sub-MHz for our operating condition. The 1 dB compression point occurs with an input power of -57 dBm.

Using the the hot ($T_H = 3.38 \pm 0.05$ K) and cold ($T_C = 20 \pm 1$ mK) loads within our setup, we calibrate the output power measured by the spectrum analyser, $P_{H,C}(\nu)$, to the noise temperature at the switch immediately following the TWPA as a function of frequency. This defines a slope, $m = (P_H - P_C)/(T_H - T_C)$, and intercept, $y_0 = P_C - mT_C$, which can be used for a y-factor noise calibration where the power measured by the spectrum analyser is referred to a noise temperature at the output of the TWPA, $T = (P-y_0)/m$. At low temperatures this relation will not be linear due to quantum mechanical effects,\cite{1928NyquistNoise} so we linearize it by using the standard technique of working in units of quanta, $n = (1/2) \coth{(h\nu/2kT)}$ instead.\cite{Caves}

By measuring the output power with the TWPA on $(P_\text{on})$ and off $(P_\text{off})$, we readily obtain the noise, in units of quanta for both cases:
\begin{equation}
\begin{aligned}
    N_\text{on}(\nu) &= \frac{P_\text{on} - y_0}{m} \left(\frac{1}{G(\nu)} \right)\left(\frac{k}{h\nu} \right) \\
    N_\text{off}(\nu) &= \frac{P_\text{off} - y_0}{m} \left(\frac{k}{h\nu} \right) 
\end{aligned}
\end{equation}
Our choice of placing the hot and cold loads at the output of the TWPA necessitates the division by the TWPA gain, $G(\nu)$, to refer this noise to the conventional reference at the input of the device. This procedure means we also avoid any complications associated with input noise at the idler frequencies that are incurred by the traditional configuration of a variable temperature load at the device input.\cite{GaoNoise}

As expected, the amplifier-off measurement, $P_\text{off}$, is identical to the cold-load, $P_C$, showing we have sufficiently attenuated the higher-temperature noise along the input channels. In practice, a second measurement of $P_\text{off}$ provides a better calibration than $P_C$ due to system drifts. Using the cold switch applies a non-negligible amount of heat to the measurement system, which then returns to the base temperature at 20 mK over the course of an hour. During that time, the combination of post-amplifiers and spectrum analyzer will drift at an average rate of 0.01 dB per 100 minutes. To minimize these drifts, we choose to use a second independent measurement of the system in the $P_\text{off}$ configuration as our 20 mK reference. Using the proper termination produces qualitatively identical results that are systematically offset by a frequency-independent level on the scale of 0.2 quanta up in an arbitrary direction depending on the system drift.

\begin{figure}[!htp]
    \centering
    \includegraphics[width=0.48\textwidth]{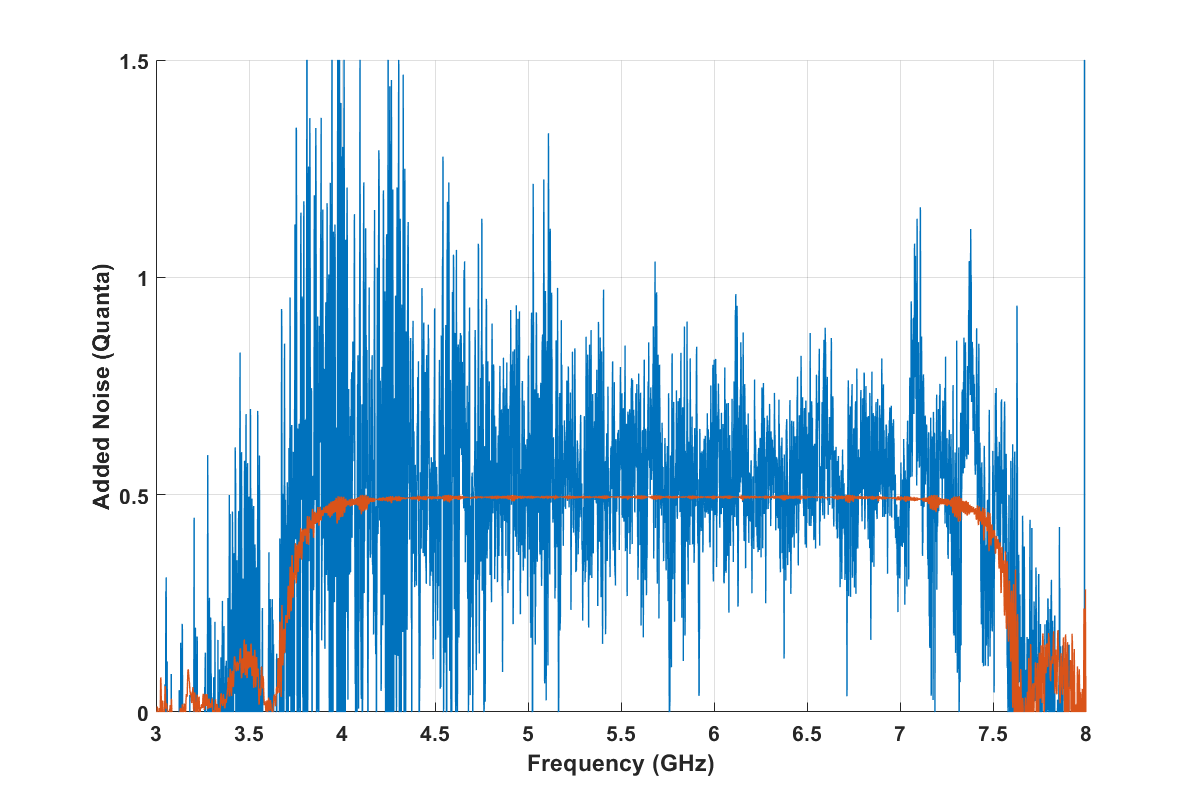}
    \caption{The measured added noise in units of quanta for the parametric amplifier operated as described in Figure \ref{fig:AddedNoiseGain} (blue) compared to the theoretical quantum limit (red).}
    \label{fig:AddedNoise}
\end{figure}

With these considerations in mind, the TWPA added noise referred to the input of the device is thus simply the difference between the on and off cases, $A(\nu) = N_\text{on}(\nu) - N_\text{off}(\nu)$. The result, shown in Figure \ref{fig:AddedNoise} demonstrates a remarkable correspondence to the well-known theoretical quantum-limit of the added noise of a phase-insensitive amplifier\cite{Caves}
\begin{equation}
A = \frac{1}{2}\left(1 - \frac{1}{G}\right)
\end{equation}
calculated using the gain in Figure \ref{fig:AddedNoiseGain}. As expected, the added noise drops to zero in regions where the amplifier does not produce gain and rises to the half-quantum limit in the regions with high gain. The lower spread in the data at high frequencies results from the increased difference between $P_H$ and $P_C$ at those frequencies, allowing for a more accurate calibration of the TWPA noise. The sinusoidal structures arise from the change in the internal ripple structure of the $S_{21}$ as a result of changing the cold switch between the three configurations used in the measurement. This noise performance coupled with the 1 dB saturation indicates the amplifier can operate with over 100 dB off dynamic range. 

A notable exception to the half-quanta standard quantum limit of added noise occurs at half the pump frequency of a three-wave-mixing TWPA, $\omega_p/2 = \omega_s = \omega_i$, where the amplifier will preferentially amplify one quadrature of the incident signal while deamplifying the other. In this phase-sensitive mode, an ideal amplifier can add no noise to the amplified quadrature while reducing the input noise in the squeezed quadrature beneath the level of vacuum fluctuations.\cite{Caves} A minor adjustment to our measurement setup allows us to measure the level of vacuum noise squeezing following the methodology of such previous experiments using JPAs.\cite{Castellanos_Beltran_2008} 

To maintain a constant relative phase between the pump and signal, both tones are provided by the same synthesizer (doubled for the pump) with a variable phase shifter and attenuator allowing for measurement of both quadratures and variable gain. The spectrum analyzer is replaced by an IQ mixer driven by the same synthesizer and allows us to measure the power in each quadrature through the noise spectral density of each channel (see Appendix B for further detail). 

\begin{figure}[!htp]
    \centering
    \includegraphics[width=0.48\textwidth]{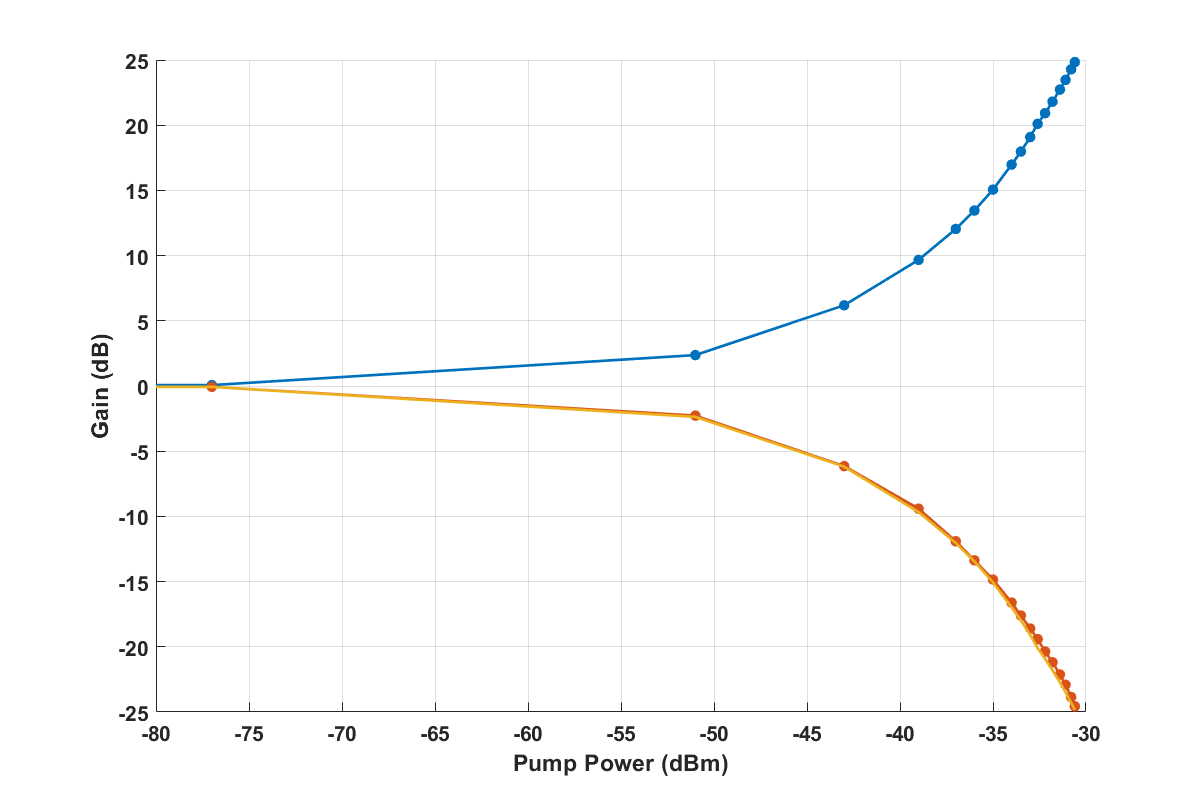}
    \caption{The measured gain at $\omega_p/2 = 5.6565$GHz in the amplified quadrature \textit{(blue)}, squeezed quadrature \textit{(red)}, and the expected squeezing level based on $G_\text{a} G_\text{sq} = 1$ \textit{(yellow)} plotted as a function of pump attenuation relative to the maximum applied pump power.}
    \label{fig:SqueezingGain}
\end{figure}

In order to optimize the measurement sensitivity, we adjust the pump tone to 11.313GHz to maximize the local frequency variations within our setup for the readout signal at $\omega_p/2$. The gain for the amplified quadrature should be identical to the inverse gain of the squeezed quadrature, i.e. $G_\text{a}G_\text{sq} = 1$, in agreement with our experimental observations (Figure \ref{fig:SqueezingGain}). 

\begin{figure}[!htp]
    \centering
    \includegraphics[width=0.48\textwidth]{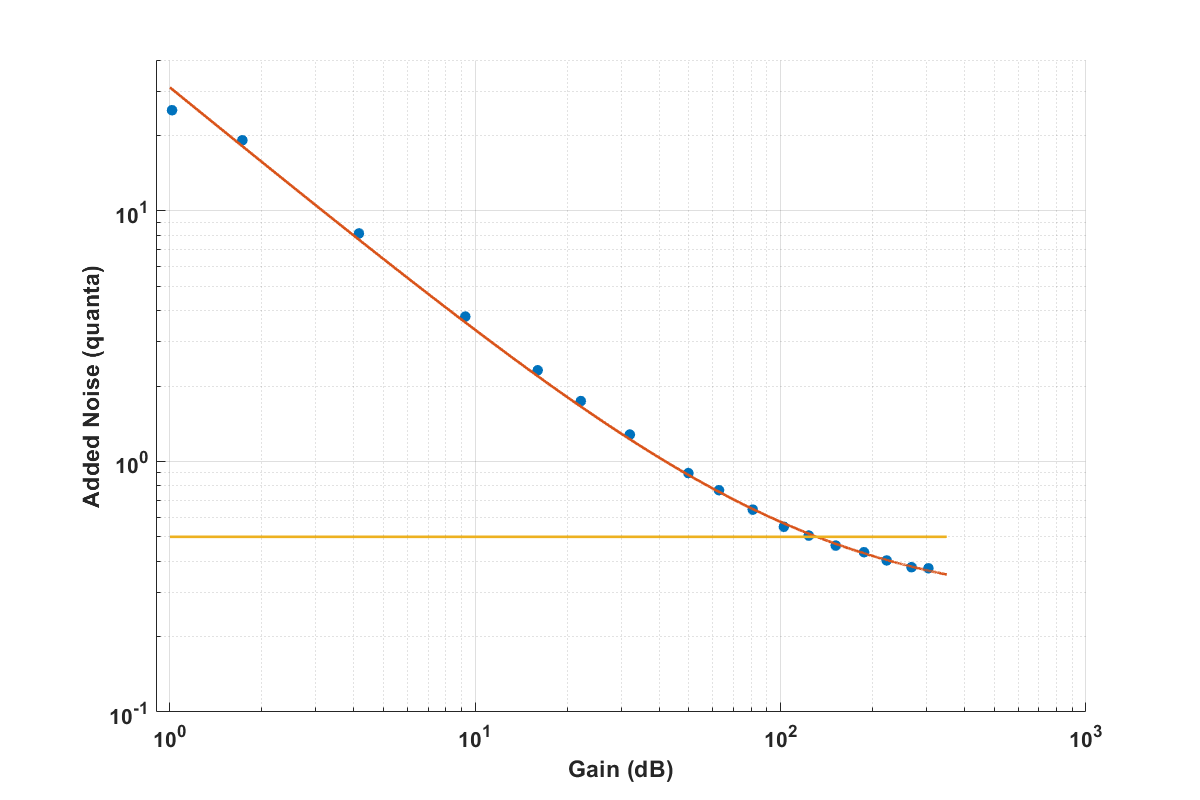}
    \caption{The measured added noise of our amplifier system referred to the input of the TWPA \textit{(blue)}, the best fit of $N_\text{HEMT} = 30.81$ and $N_\text{a} = 0.27$ \textit{(red)}, and the $0.5$ quanta level of vacuum fluctuations \textit{(blue)} plotted against the amplified quadrature TWPA gain.}
    \label{fig:Nsys}
\end{figure}

In the approximation that the system noise is dominated by the combination of the TWPA and HEMT,\cite{FRIIS} the added noise of our readout amplifier system for the amplified quadrature will be given by
\begin{equation}
N_\text{sys} = N_\text{a} + \frac{N_\text{HEMT}}{G_\text{a}}
\end{equation}
where $N_\text{a}$ is the TWPA added noise and $N_\text{HEMT}$ is the HEMT noise propagated back to the input of the TWPA (through an experimentally measured 4 dB of attenuation within our system).\cite{Castellanos_Beltran_2008} Performing a y-factor measurement of our noise using the same methodology as the previous section, we obtain a the total amplifier system added noise ($N_\text{sys}$) and can fit for the TWPA contribution (see Figure \ref{fig:Nsys}). When the TWPA is operated with high gain, the system noise drops beneath the phase-insensitive standard quantum limit for this quadrature.

\begin{figure}[!htp]
    \centering
    \includegraphics[width=0.48\textwidth]{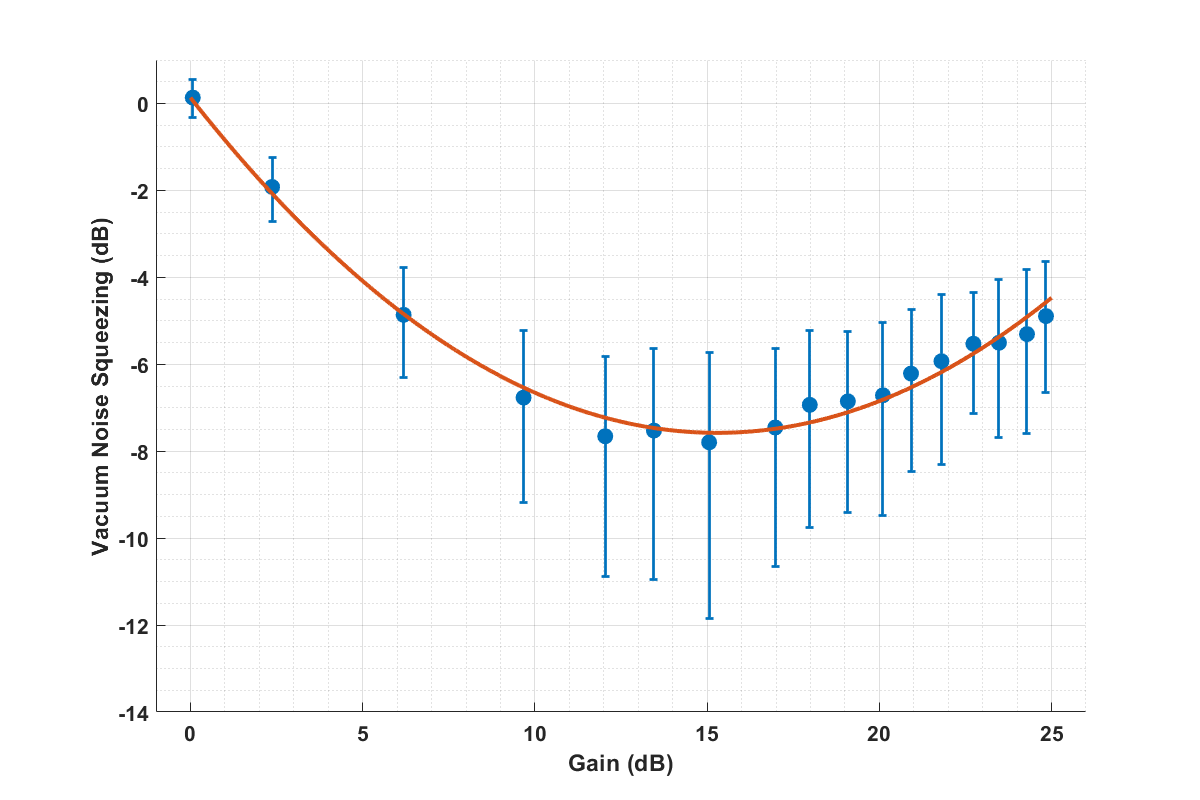}
    \caption{The vacuum squeezing level obtained by the TWPA in the squeezed quadrature plotted against the amplified quadrature gain.}
    \label{fig:SqueezingLevel}
\end{figure}

In the squeezed quadrature, the TWPA input noise will be reduced as it propagates through the device, resulting in a squeezed vacuum state. The measured output noise will be a function of the system input 20mK noise ($N_\text{mK} \approx 0.5$), TWPA squeezing factor ($G_\text{sq}$), TWPA to HEMT attenuation ($A$), and HEMT noise ($N_\text{HEMT}$):
\begin{equation}
    N_\text{sq} = \left(\frac{N_\text{mK}}{G_\text{sq}} + N_\text{pa}\right) A + N_\text{mK}(1 - A) + N_\text{HEMT}.
\end{equation}
Taking the ratio between the TWPA on and off cases, we can use the reduction in measured noise to solve for $G_\text{sq}$ and calculate the vacuum noise squeezing level we obtain immediately following the TWPA. The result is shown in Figure \ref{fig:SqueezingLevel} demonstrating a maximum of 8 dB vacuum noise squeezing. In the ideal noise-less limit, the squeezing level should be perfectly linear with the amplified quadrature gain. The deviation corresponds to a maximum TWPA noise of $N_\text{sq} = 0.16$ quanta in the squeezed quadrature.

In conclusion, we have experimentally demonstrated a working traveling-wave superconducting parametric amplifier based on the nonlinear kinetic inductance of a NbTiN microstrip. This amplifier produces roughly 20 dB gain across several GHz of bandwidth centered around 5.65 GHz. We further demonstrate a noise performance with excellent correspondence to the standard quantum limit. A further investigation of the noise at half the pump frequency shows both improved noise performance in the amplified quadrature and substantial vacuum noise squeezing in the other. A TWPA operating in this regime can result in significant improvements to the sensitivity of microwave experiments.\cite{ramanathan2022wideband}

\bibliography{refs}

\clearpage

\appendix

\section{Pump Heating Effect}

Since the gain of a three-wave mixing amplifier depends on the product $I_\text{DC}I_p$,\cite{Shu_2021} optimal gain is achieved by setting $I_\text{DC}$ to half of $I_c$ and raising the pump power to the maximum level before breaking superconductivity. A similar level of gain can be achieved by marginally lowering either of these currents and raising the other to compensate. For this device, we would set $I_\text{DC} = 0.3$mA and $I_p = -15.7$ dBm for optimal gain, $0.26$mA lower and $1.5$ dB higher than the operating condition for the gain shown in Figure \ref{fig:AddedNoiseGain} respectively. 
\begin{figure}[!htp]
    \centering
    \includegraphics[width=0.48\textwidth]{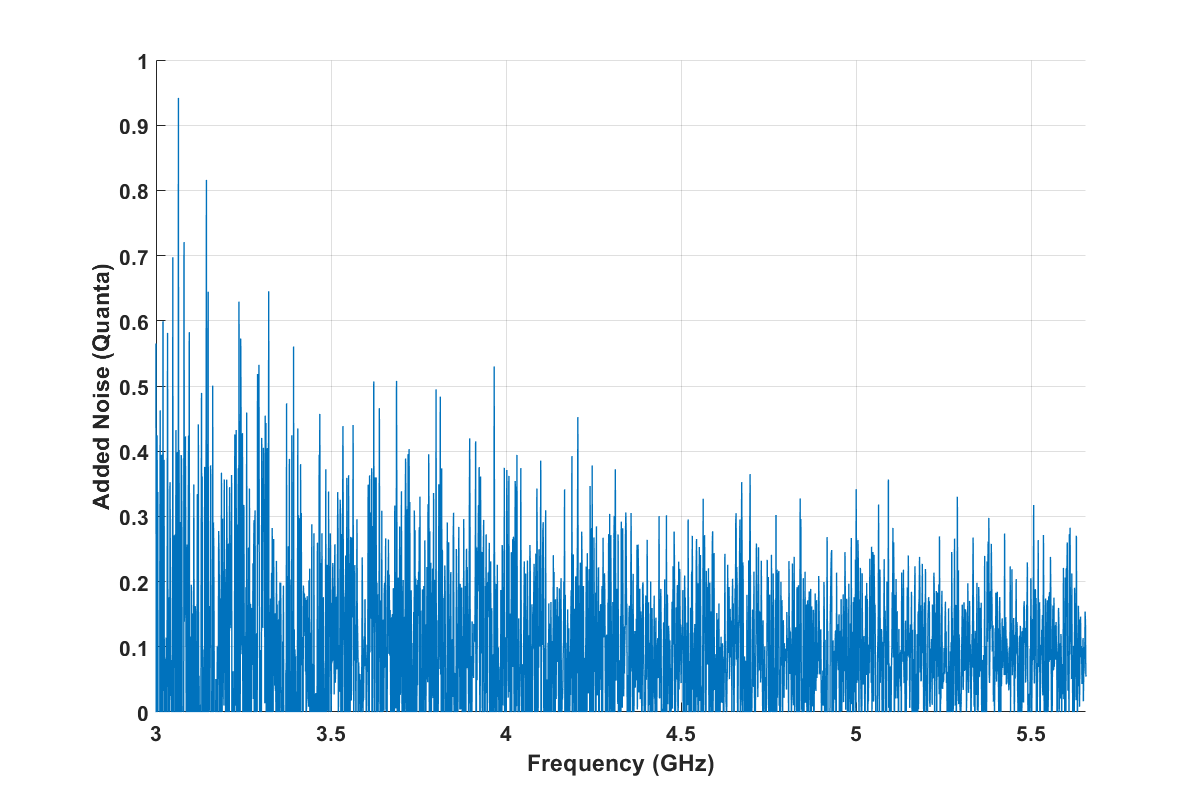}
    \caption{The measured added noise from an applied $I_p = -15.7$dBm at the output of the signal generator.}
    \label{fig:PumpNoise}
\end{figure}

We choose to operate in this higher-DC, lower-pump regime due to a heating effect in our mK-stage setup in the presence of a strong applied pump tone. Turning off the DC current, and repeating the noise measurement with only an applied pump tone results in a frequency-independent increase in the measured noise as shown in Figure \ref{fig:PumpNoise}. On average, we note a $0.077$ quanta excess in noise for a $-15.7$ dBm pump which steadily decreases as the pump power is lowered. It is reasonable to assume that this effect is in large part responsible for the small amount of excess noise measured in our vacuum squeezing measurements. 

\vspace{20pt}
\newpage

\section{Noise Squeezing Measurement Setup}

\begin{figure}[!htp]
    \centering
    \includegraphics[width=0.48\textwidth]{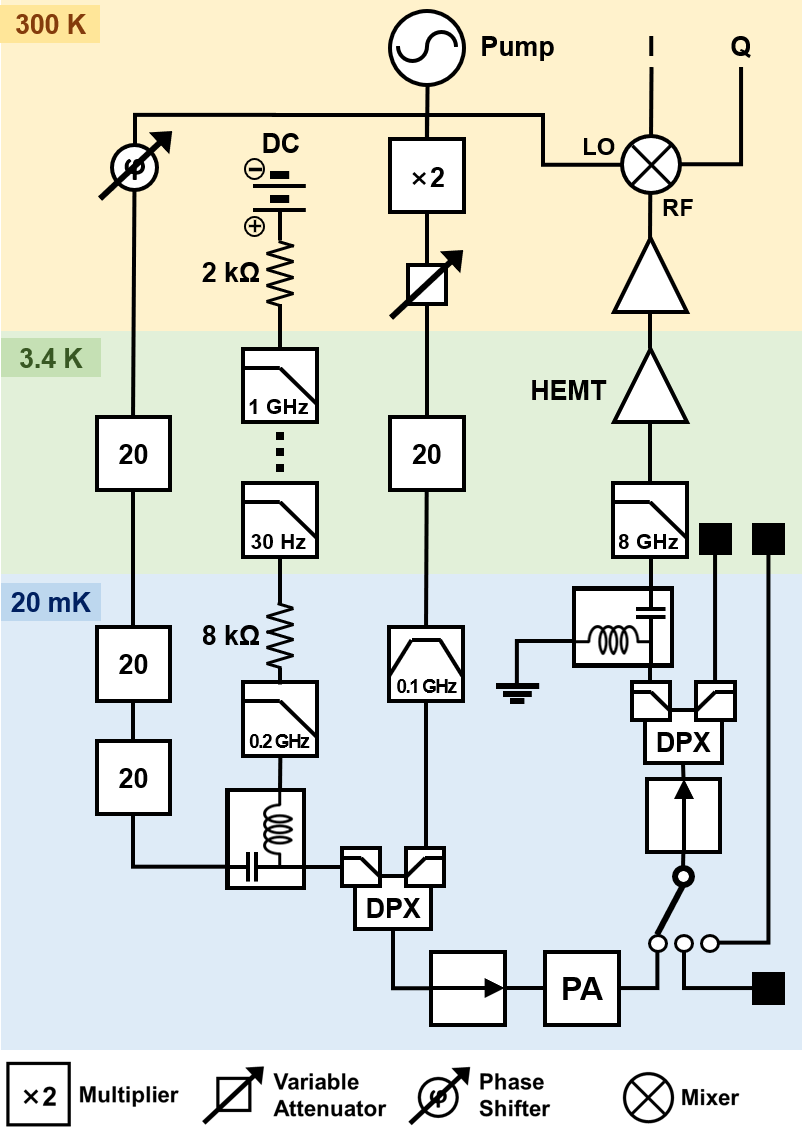}
    \caption{The altered room-temperature setup used for degenerate vacuum squeezing measurements.}
    \label{fig:SqueezingSetup}
\end{figure}

The modified room-temperature experimental setup for degenerate vacuum squeezing measurements is shown in Figure \ref{fig:SqueezingSetup}. The signal generator is set to produce a tone at $\omega_p/2$ which is split into a three components. The first (left) is heavily attenuated (not shown) and sent through a variable phase shifter to act as a test tone for measuring the degenerate gain for both quadratures. The second (center) is multiplied back to $\omega_p$ and serves as the TWPA pump using a digital variable attenuator to control its power. The third (right) drives the LO for a mixer to convert the output power in both quadratures $\omega_p/2$ to an $I$ and $Q$ stream that is then amplified, low-pass-filtered, and read out by a DAQ.

\end{document}